\documentclass[aps,prl,twocolumn,superscriptaddress,showpacs]{revtex4-1}
\usepackage[version=3]{mhchem}
\usepackage[english]{babel}
\usepackage{dcolumn}
\usepackage{graphicx}
\usepackage{amsmath}
\usepackage{longtable}
\usepackage{bigstrut}

\begin{document}


\title{Static trapping of polar molecules in a traveling wave decelerator}


\author{Marina Quintero-P\'{e}rez}
\affiliation{LaserLaB, Department of Physics and Astronomy, VU University Amsterdam, 
De Boelelaan 1081, 1081 HV Amsterdam, The Netherlands}
\author{Paul Jansen}
\affiliation{LaserLaB, Department of Physics and Astronomy, VU University Amsterdam, 
De Boelelaan 1081, 1081 HV Amsterdam, The Netherlands}
\author{Thomas E. Wall}
\affiliation{LaserLaB, Department of Physics and Astronomy, VU University Amsterdam, 
De Boelelaan 1081, 1081 HV Amsterdam, The Netherlands}
\author{Joost~E.~van~den~Berg}
\affiliation{University of Groningen, KVI, Zernikelaan 25, 9747 AA, Groningen, The Netherlands}
\author{Steven Hoekstra}
\affiliation{University of Groningen, KVI, Zernikelaan 25, 9747 AA, Groningen, The Netherlands}
\author{Hendrick L. Bethlem}
\affiliation{LaserLaB, Department of Physics and Astronomy, VU University Amsterdam,
De Boelelaan 1081, 1081 HV Amsterdam, The Netherlands} 


\date{\today}

\begin{abstract}
We present experiments on decelerating and trapping ammonia molecules using a combination of a Stark decelerator and a traveling wave decelerator. In the traveling wave decelerator a moving potential is created by a series of ring-shaped electrodes to which oscillating high voltages are applied. By lowering the frequency of the applied voltages, the molecules confined in the moving trap are decelerated and brought to a standstill. As the molecules are confined in a \emph{true} 3D well, this kind of deceleration has practically no losses, resulting in a great improvement on the usual Stark deceleration techniques. The necessary voltages are generated by amplifying the output of an arbitrary wave generator using fast HV-amplifiers, giving us great control over the trapped molecules. We illustrate this by experiments in which we adiabatically
cool trapped NH$_{3}$ and ND$_{3}$ molecules and resonantly excite their motion.
\end{abstract}

\maketitle
\footnotetext{\ddag~These authors contributed equally to this work}

Cold molecules offer fascinating prospects for precision tests of fundamental physics theories, cold chemistry and quantum computation (for recent review papers see~\cite{Carr:NJP2009,Hogan2011,vandeMeerakker:ChemRev2012,Narevicius:ChemRev2012}). One of the methods that has been successful in producing samples of cold and trapped molecules is Stark deceleration. This technique uses a series of strong electric fields that are switched at the appropriate times to lower the velocity of a beam of molecules in a stepwise fashion. In conventional Stark decelerators the principle of phase stability is employed, which ensures that molecules experience an \emph{effective} 3D potential well that keeps them in a compact packet during the deceleration process~\cite{Bethlem:PRL2000,Bethlem:PRA2002,vandeMeerakker:ChemRev2012}. Whereas the approximations used to derive this effective well are valid at high velocity, they no longer hold at low velocity, when the characteristic wavelength of the longitudinal and transverse motion of the molecules becomes comparable to the periodicity of the decelerator. As a result of this breakdown, the number of molecules in the packet as well at its phase-space density decrease at low velocities~\cite{Jansen:Thesis}. Further losses take place when the decelerated molecules are loaded into a trap~\cite{Bethlem:PRA2002,Sawyer:PRL2007,Gilijamse:EPJD2010}.

Expanding on earlier work on chip-based Stark decelerators~\cite{Meek:PRL2008}, Osterwalder and co-workers~\cite{Osterwalder2010,Meek2011} decelerated CO molecules from 288\,m/s to 144\,m/s using a series of ring electrodes to which a sinusoidal voltage was applied. In this traveling wave decelerator decelerator the molecules experience a \emph{genuine} rather than an \emph{effective} potential well that moves continuously along with the molecules. The molecules are decelerated by slowly decreasing the velocity of the moving potential well by lowering the frequency of the applied voltages. Besides avoiding losses at low velocities, a traveling wave decelerator has the advantage that molecules can be brought to a complete standstill and trapped without the need to load them into a separate electrode geometry. Furthermore, as the molecules are always close to the zero-point of the electric field, they can be decelerated in states that become high-field seeking at relatively low electric fields. This makes it possible to decelerate heavy molecules such as YbF~\cite{Bulleid:PRA2012} and SrF~\cite{Berg:EPJD2012}. 

In this Letter, we use a traveling wave decelerator to decelerate and trap ammonia molecules. Our main motivation for this research is the possibility to use the traveling wave decelerator as a source of cold molecules for a molecular fountain~\cite{Bethlem:EPJST2008}. Previous attempts to create a fountain using a Stark decelerator were unsuccessful due to losses at low velocities and a complex lens-system for cooling and collimating the slow beam~\cite{Jansen:Thesis}. A traveling wave decelerator should solve both of these issues.    

 \begin{figure}
 \includegraphics{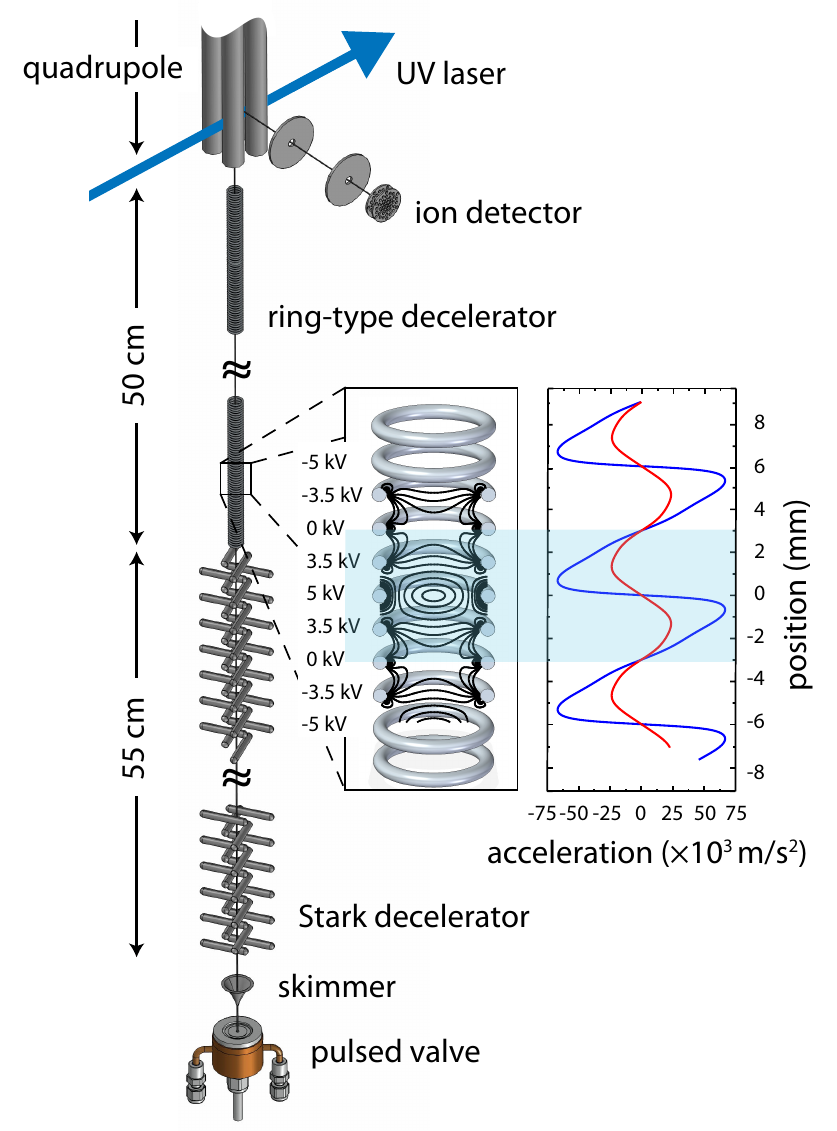}%
 \caption{\label{fig1-setup}
 Schematic view of the vertical molecular beam machine. The inset shows the electric field magnitude inside the ring-type decelerator with the voltages on the different electrodes as indicated. The electric field magnitude is shown in steps of 2.5~kV/cm. In the rightmost panel the longitudinal acceleration on NH$_{3}$ (dashed red curve) and ND$_{3}$ (solid blue curve) molecules is shown along the molecular beam axis.}
 \end{figure}

Fig.~\ref{fig1-setup} shows a schematic view of the vertical molecular beam machine. A supersonic beam of ammonia molecules is decelerated to around 100~m/s using a conventional decelerator consisting of a series of 100 electrode pairs to which voltages of $\pm$10~kV are applied (for details of the beam machine and Stark deceleration of ammonia, see~\cite{Bethlem:PRA2002,Bethlem:EPJST2008,Quintero-Perez:PCCP2012}). A traveling wave decelerator is mounted 24~mm above the last electrode pair. The design and dimensions of the decelerator are copied from Osterwalder and co-workers~\cite{Osterwalder2010,Meek2011}. The decelerator consists of 336 ring electrodes with an inner diameter of 4\,mm which are attached to one of eight supporting bars. Consecutive rings are separated by 1.5\,mm (center to center), resulting in a periodic length, $L$, of 12\,mm. The voltages applied to the eight support bars are generated by amplifying the output of an arbitrary wave generator (Wuntronic DA8150) using eight fast HV-amplifiers (Trek 5/80) up to $\pm$5\,kV. A 50\,cm long quadrupole is mounted 20\,mm above the traveling wave decelerator. In the experiments described in this Letter, the quadrupole is used only to extract ions that are produced by a focused UV laser that crosses the molecular beam 40\,mm above the last ring electrode. The ions are counted by an ion detector. 

At any moment in time, the voltages applied to successive ring electrodes follow a sinusoidal pattern in $z$, where $z$ is the position along the beam axis. These voltages create a minimum of electric field every 6\,mm, representing a true 3D trap for weak-field seeking molecules. By modulating these voltages sinusoidally in time the traps can be moved along the decelerator. The velocity of the trap is given by $v_{z}(t) = f(t)L$, with $f$ being the modulation frequency. A frequency that is constant in time results in a trap that moves with a constant positive velocity along the cylindrical axis. A constant acceleration or deceleration can be achieved by applying a linear chirp to the frequency~\cite{Osterwalder2010,Meek2011}. 

The inset of Fig.~\ref{fig1-setup} shows the electric field magnitude inside the ring decelerator with the voltages on the different electrodes as indicated. The electric field near the center has a quadrupolar symmetry, i.e., the electric field magnitude increases linearly away from the center. It is essential to the operation of the decelerator that the trapping potential maintains a constant shape and depth while it is moved. In the chosen geometry, the electric field gradients in the bottom of the well, as well as the trap depth in the longitudinal direction are nearly independent of the position of the trap minimum. The trap depth in the transverse direction, however, is 40\% deeper when the trap minimum is located in the plane of a ring compared to the situation when the trap minimum is located in the middle between two rings. 

The panel on the right hand side of Fig.~\ref{fig1-setup} shows the acceleration experienced by NH$_{3}$ (dashed red curve) and ND$_{3}$ (solid blue curve) molecules as a function of $z$. The inversion splitting in NH$_{3}$ is 23.8\,GHz, while it is only 1.4\,GHz in ND$_{3}$. As a result, the Stark effect in NH$_{3}$ is quadratic up to electric fields of 20\,kV/cm and  NH$_{3}$ molecules experience an acceleration that increases linearly with the distance from the trap center. The Stark effect in ND$_{3}$, on the other hand,  becomes linear at much smaller fields and the ND$_{3}$ molecules experience a linearly increasing acceleration close to the trap center only. As a result of the different shape (and magnitude) of the force, the dynamics of these two ammonia isotopologues in the decelerator is markedly different, as will be seen in the experimental results. 

 \begin{figure}
 \includegraphics{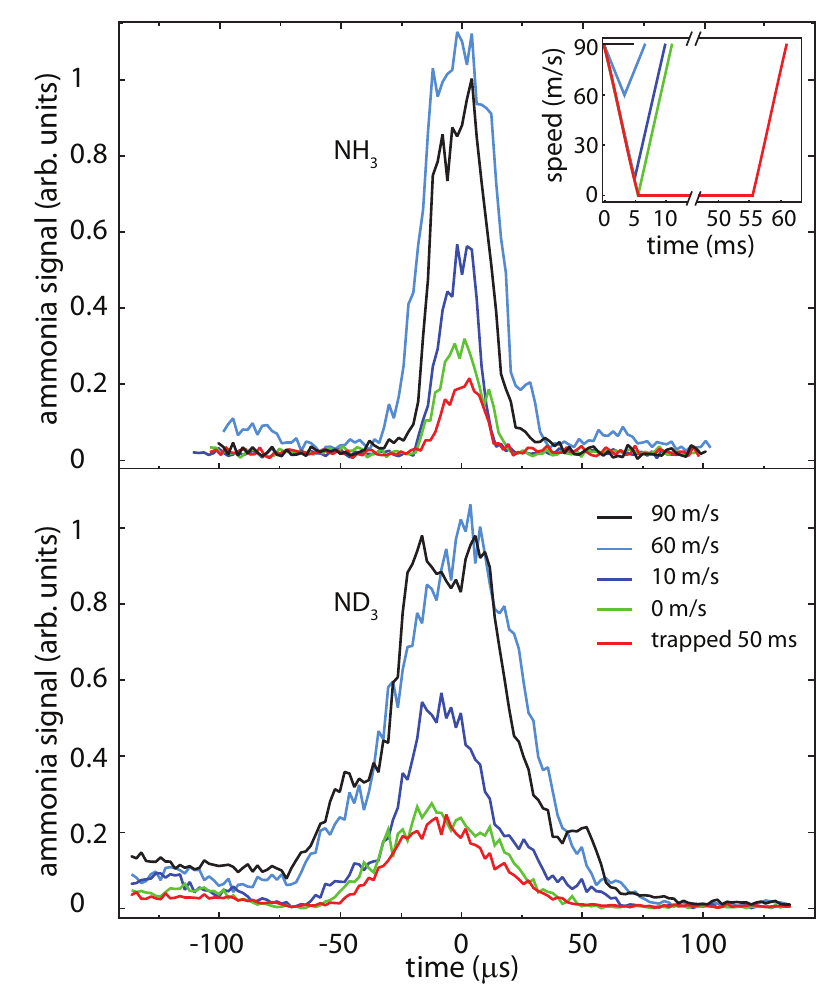}%
 \caption{\label{fig2-TOFs}
Measured time-of-flight (TOF) profiles for NH$_{3}$ (upper panel) and ND$_{3}$ (lower panel) when molecules are guided at 90\,m/s (black curve), decelerated to 60, 30 and 0\,m/s (light blue, dark blue and green, respectively) and trapped for 50\,ms (red curve) before being accelerated back to 90\,m/s and detected. The time of flight traces have been centered around the expected arrival time. The inset shows the velocity of the electric field minimum as a function of time for the different recorded TOF profiles.}
 \end{figure}

Fig.~\ref{fig2-TOFs}  shows the density of NH$_{3}$ molecules (upper panel) and ND$_{3}$ molecules (lower panel) above the decelerator as a function of time when different waveforms are applied to the ring decelerator as shown in the inset. In all cases, a packet of molecules is decelerated to 90\,m/s using the conventional Stark decelerator and injected into the traveling wave decelerator. The horizontal axis is always centered around the expected arrival time of the molecules. The black curves show the density when a sinusoidal voltage with an amplitude of 5\,kV and a (constant) frequency of 7.5\,kHz is applied to the ring decelerator. With these settings, molecules traveling at 90\,m/s are guided through the ring decelerator, resulting in a Gaussian distribution around the origin. The width of the time-of-flight (TOF) profile mainly reflects the velocity spread of the guided molecules. As for ND$_{3}$ the moving trap is deeper than for NH$_{3}$, its TOF profile is accordingly wider. Wings are observed to earlier and later arrival times, which are attributed to molecules that are being trapped in a minimum of the electric field that is 12\,mm above or below the synchronous one.  

The light blue, dark blue and green curves in Fig.~\ref{fig2-TOFs} show the recorded TOF traces when the frequency is first decreased linearly with time to a value of 5, 0.83 and 0\,kHz, respectively, and subsequently increased to its original value. With these settings, molecules are decelerated to 60, 10 and 0\,m/s  before being accelerated back to 90\,m/s, using accelerations of $\pm$9.2, $\pm$16.4, and $\pm$16.6\,10$^{3}$\,m/s$^2$, respectively. 

The observed NH$_{3}$ and ND$_{3}$ density decrease at lower velocities is larger than expected from simulations. We attribute the loss mainly to mechanical misalignments that lead to parametric amplification of the motion of the trapped molecules at low velocities. On inspection, it was noticed that one of the suspension bars was slightly displaced from its original position,
which must have happened when the decelerator was placed in the vacuum chamber. Another loss mechanism comes from the fact that the phase space distribution of the packet exiting the conventional Stark decelerator is not perfectly matched to the acceptance of the traveling wave decelerator. As a result, the trapped packet will perform a (damped) breathing motion. This oscillation explains why the observed TOF profile for deceleration to 60\,m/s is wider and more intense than the TOF profile for guided molecules. Neither loss mechanism is fundamental, and we believe they can be eliminated in future work. 

The red curves in Fig.~\ref{fig2-TOFs} are recorded when, after the frequency of the applied voltages is decreased to 0\,kHz, the voltages are kept at constant values for 50\,ms before the frequency is increased to its original value of 7.5\,kHz. The observed TOFs are almost identical to the ones recorded when the frequencies are immediately returned to their original value. This measurement demonstrates that molecules can be trapped in the laboratory frame without further losses.

 \begin{figure}
 \includegraphics{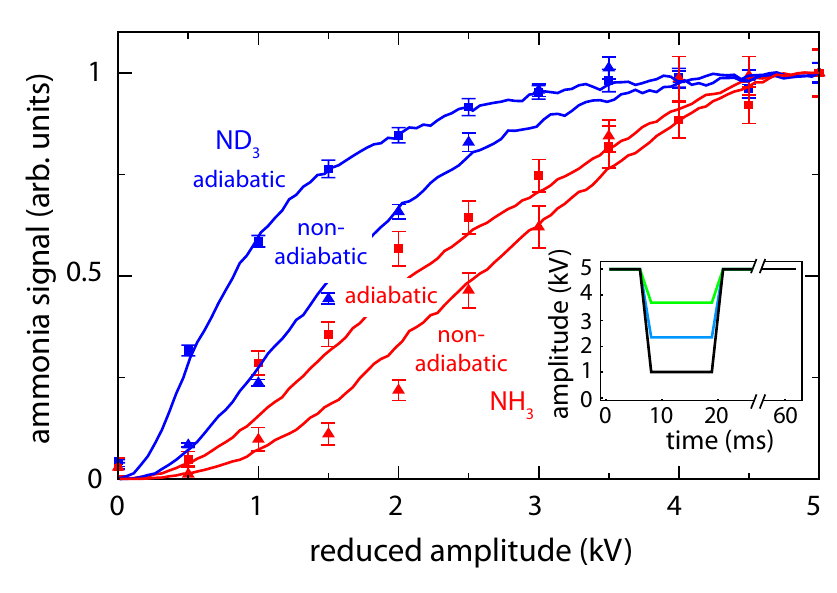}%
 \caption{\label{fig3-adiabatic}
Trapped NH$_{3}$ (red data points) and ND$_{3}$ (bue data points) signal as a function of the amplitude of the waveform when, after deceleration, the trap is slowly (squares) or abruptly (triangles) reduced. The solid lines are simulations that assume an initial temperature of 14 and 30\,mK for NH$_{3}$ and ND$_{3}$, respectively. The inset shows the amplitude of a number of typical waveforms as a function of time.}
 \end{figure}

As the voltages applied to the decelerator are generated by amplifying the output of an arbitrary wave generator using fast HV-amplifiers, we can change the depth (and shape) of the trap at will. This is illustrated in the measurements shown in Fig.~\ref{fig3-adiabatic}. In these measurements NH$_{3}$ (red squares) and ND$_{3}$ (blue squares) molecules are again decelerated, trapped for a period of over 50\,ms, and subsequently accelerated and detected. In this case, however, while the molecules are trapped, the voltages applied to the decelerator are ramped down (in 2\,ms for ND$_{3}$  and 10\,ms for NH$_{3}$), kept at a lower value for 10\,ms and then ramped up to 5\,kV. Typical waveform amplitudes as a function of time are shown in the inset. Lowering the voltages of the trap has two effects: (i) the trap frequency is lowered, adiabatically cooling the molecules; (ii) the trap depth is reduced, allowing the hottest molecules to escape the trap. The solid lines show the result of a simulation assuming an (initial) temperature of 14\,mK for NH$_{3}$ and 30\,mK for  ND$_{3}$. Note that we use temperature here only as a convenient means to describe the distribution; the densities are too low to have thermalization on the timescales of the experiment. For comparison, the blue triangles show measurements when the trap voltages are abruptly (10\,$\mu$s) lowered. In this case no adiabatic cooling occurs and as a result the signal decreases more rapidly when the voltages are ramped to lower voltages. The solid lines again show the result of a simulation. It is seen that, both in the adiabatic and in the nonadiabatic case, the NH$_{3}$ signal drops more rapidly than the ND$_{3}$ signal. This is mainly because the ratio of the temperature of the molecules to the trap depth at 5\,kV is larger for NH$_{3}$ than for ND$_{3}$, i.e., the NH$_{3}$ molecules initially fill the trap almost completely while the ND$_{3}$ molecules only occupy a fraction of the trap. We have also measured the number of molecules that remain trapped at 2\,kV as a function of the time used for lowering the voltages (not shown). These measurements confirm that the ND$_{3}$ and NH$_{3}$ molecules follow the trap adiabatically when the ramping times are longer than 1\,ms. 

 \begin{figure}
 \includegraphics{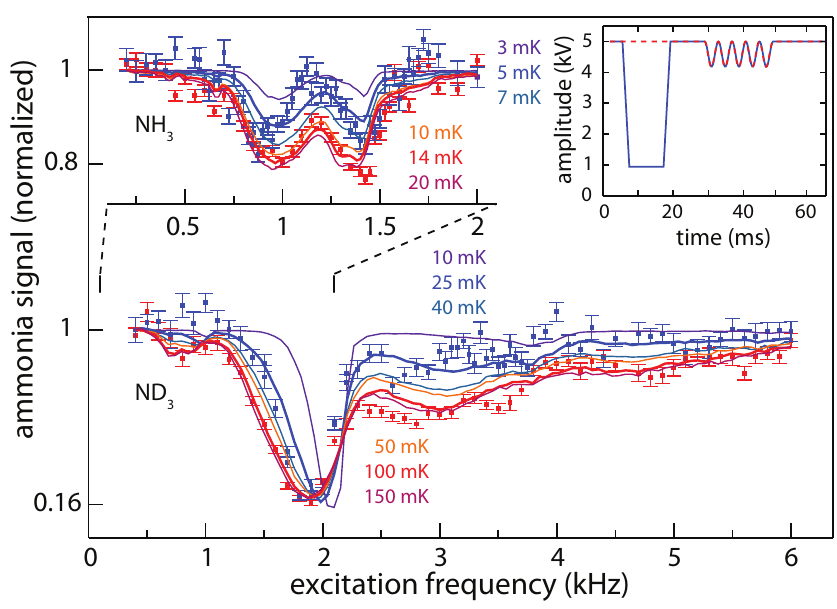}%
 \caption{\label{fig4-Resonances}
Fractional trap loss as a function of excitation frequency for NH$_{3}$ (upper, offset for clarity) and ND$_{3}$ (lower) with (blue data points) and without (red data points) reducing the trap depth. In the inset, the amplitude of the waveform is shown as a function of time. The amplitude is modulated by 400\,V for NH$_{3}$, and by 800\,V for ND$_{3}$. The solid curves result from 3D simulations that assume a truncated thermal distribution with temperatures as indicated in the figure.}
 \end{figure}

Fig.~\ref{fig4-Resonances} shows the normalized signal of trapped NH$_{3}$ (upper panel) and ND$_{3}$ molecules (lower panel) when the amplitude of the trapping voltages is modulated. If the frequency of the modulation matches a characteristic frequency of a trapped molecule, the amplitude of the motion of the molecule inside the trap is increased~\cite{LandauMechanics}. As a result, the density is decreased. The excitation scans are recorded in two situations. In the first case (red data points), molecules are decelerated and trapped for 30\,ms before the excitation voltages are applied, after which they are accelerated and detected. In the second case (blue data points), the trap voltages are adiabatically lowered for a period of 10\,ms before the excitation voltages are applied, allowing the hottest molecules to escape the trap. The amplitude of the waveform for both cases is shown in the inset. 

For NH$_{3}$, two strong resonances are seen around 0.9 and 1.4\,kHz, assigned to $2f_{r}$ and $2f_{z}$, respectively, where $f_{r}$ and $f_{z}$ are the radial and longitudinal trap frequency, respectively. For ND$_{3}$, a strong resonance is observed around 1.9\,kHz and a weaker resonance around 3\,kHz. These resonances are assigned to $2f_{r}$ and $f_{r}+f_{z}$, respectively. Due to the anharmonicity of the trap, the width of the observed resonances depends strongly on temperature. The solid curves show simulations that assume a thermal distribution with temperatures as indicated in the figure. Note that for the ND$_{3}$ measurements presented in Fig.~\ref{fig4-Resonances}, the Stark decelerator was operated at a lower phase angle, which explains why the temperature determined from these measurements is much larger than that found from the measurements presented in Fig.~\ref{fig3-adiabatic}. These data show the great control allowed by the decelerator, using the molecules to map out the trapping potentials, and using this mapping as direct evidence of phase-space manipulation in the trap.

In conclusion, we have decelerated and trapped ammonia  (NH$_{3}$ and ND$_{3}$) molecules using a combination of a Stark decelerator and a traveling wave decelerator. We take advantage of the strong acceleration of a conventional Stark decelerator to decelerate molecules to around 100\,m/s, removing 90\% of their kinetic energy. We then exploit the stability of a traveling wave decelerator to bring molecules to a standstill inside the decelerator. In this way, the voltages are varied over a frequency range that is covered by commercially available HV-amplifiers. We demonstrate that this setup can be used to change the position, shape and depth of the trap at will, giving us unique control over the trapped molecules. This control will greatly facilitate the implementation of schemes to further cool the molecules such as sisyphus cooling \cite{Zeppenfeld:Nature2012} or evaporative cooling \cite{Stuhl:Nature2012}.       

\begin{acknowledgments}
This research has been supported by NWO via a VIDI-grant, by the ERC via a Starting Grant and by the FOM-program `Broken Mirrors \& Drifting Constants'. We acknowledge the expert technical assistance of Jacques Bouma, Joost Buijs, Leo Huisman and Imko Smid. We thank Andreas Osterwalder and Gerard Meijer for helpful discussions and Wim Ubachs for his continuing interest and support.
\end{acknowledgments}

\end{document}